\newcounter{myctr}
\def\myitem{\refstepcounter{myctr}\bibfont\noindent\ifnum\themyctr>9\else\phantom{0}\fi\hangindent17pt\themyctr.\enskip}

\documentclass{ws-ijqi}
\usepackage{hyperref}
\usepackage[super,sort,compress]{cite}
\begin{document}

\catchline{}{}{}{}{}

\title{Quantum synchronization and correlations of two qutrits in a non-Markovian bath}

\author{Jian-Song Zhang}

\address{Department of Applied Physics, East China Jiaotong University,
Nanchang 330013, People's Republic of China\\
jszhang1981@zju.edu.cn}

\maketitle

\begin{history}
\received{Day Month Year}
\revised{Day Month Year}
\end{history}

\begin{abstract}
We investigate quantum synchronization and correlations of two qutrits in one non-Markovian environment using the hierarchy equation method.
There is no direct interaction between two qutrits and each qutrit interacts with the same non-Markovian environment.
The influence of the temperature of the bath, correlation time, and coupling strength between qutrits and bath
on the quantum synchronzation and correlations of two qutrits are studied without the Markovian, Born,
and rotating wave approximations. We also discuss the influence of dissipation and dephasing on the synchronization of two qutrits.
In the presence of dissipation, the phase locking between two qutrits without any direct interaction can be
achieved when each qutrit interacts with the common bath. Two qutrits within one common bath can not be syncrhonized in the purely dephasing case.
In addition, the Arnold tongue can be significantly broadened by decreasing the correlation time of two qutrits and bath. Markovian baths
are more suitable for synchronizing qutrits than Non-Markovian baths.
\end{abstract}



\markboth{Authors' Names}
{Instructions for Typing Manuscripts (Paper's Title)}

\section{Introduction}
Synchronization which describes the adjustment of rhythms of self-sustained oscillators due to an interaction
is a fundamental phenomenon of nonlinear sciences.
This phenomenon has been observed in physical, chemical, biological, and social systems \cite{1}.
In recent years, many efforts have been devoted to extend the concept of synchronization to quantum systems
such as Van der Pol oscillators \cite{2,3,4}, atomic ensembles \cite{5,6}, trapped ions \cite{7},
and cavity optomechanics \cite{8,9,10,11}.

In general, a quantum system can be either continuous or discrete.
In the previous studies \cite{2,3,4,5,6,7,8,9,10,11},
most authors have considered the quantum synchronization of continuous-variable systems
with classical analogs since they can be described by quasiprobability distributions in phase space
such as the Wigner function. For example, in Refs. \cite{8,9,10,11}, the authors have investigated
the quantum synchronization of optomechanical systems formed by optical and mechanical modes \cite{12,13,14,15}.
The measures of complete and phase synchronization of continuous-variable quantum systems have been proposed \cite{16}.

For discrete-variable systems without classical analogue, the Pearson product-moment correlation coefficient can be
used to measure the degree the synchronization of spin systems \cite{17}.
The authors have investigated the synchronization of two qubits in a common environment using the Bloch-Redfield
master equation and found that two qubits can not be synchronized for purely dephasing case \cite{17}.
Recently, the measure of quantum synchronization using the Husimi Q representation and the
concept of spin coherent states has been suggested by Roulet and Bruder \cite{18}. This measure can be used to
study the synchronization of discrete-variable systems including qubits and qutrits.
The authors have pointed out that qubits can not be synchronized since
they lack a valid limit cycle and a spin 1 could be phase-locked to a weak external driving \cite{18}.
Later, the authors investigated the quantum synchronization and entanglement generation of two
qutrits using the Lindblad master equation \cite{19}. Very recently, the quantum synchronization of two quantum oscillators
within one common dissipative environment at zero temperature was investigated
with the help of a path integral formalism \cite{20}.

In the previous works \cite{17,18,19}, the Markovian and Born approximations were employed and the temperature of bath was assumed to be zero.
Note that the rotating wave approximation was used in the previous work \cite{20}.
Thus, the influence of the temperature of the bath or the non-Markovian effects were not taken into accounted in the above works.
In the present paper, we study the quantum synchronization and correlations of two qutrits within one common bath
using the hierarchy equation method \cite{21,22,23,24}. The two qutrits have no direct interaction.
In particular, in the derivation of hierarchy equations,
the Markovian, Born, and rotating wave approximations are not used. The hierarchy equation method is a
high-performance method and is suitable for strong- and ultrastrong-coupling systems
like chemical and biophysical systems \cite{25,26,27,28}.
Our results show that the measures of quantum synchronization and correlations could increase with the
increase of the coupling strength between each qutrit and the common bath.
The influence of the temperature of the bath depends heavily on the detuning of two qutrits.
If the detuning is much smaller than the frequecies of two qutrits, then
the maximal value of the measure of quantum synchronization increases with the increase of the temperature of the bath.
However, if the detuning is not much smaller than the frequencies of
two qutrits, the temperature of the bath could play a destructive role in the synchronization of two qutrits.
In addition, the correlation time of the qutrits and bath plays an important role in the generation of
quantum synchronization and correlations. The phase locking between two qutrits without direct interaction can
be achieved if they are put into one bath and the dissipation is taken into accounted. In particular, two
qutrits can not be synchronized in the purely dephasing case.
The Arnold tongue of synchronization and quantum correlations (measured by quantum mutual information)
can be obtained in the present model. The shape of the Arnold tongue can be adjusted by the temperature, coupling strength, and correlation time
of the system.

The organization of this paper is as follows. In Sec. II, we introduce the model and
the hierarchy equation method. In Sec. III, we briefly review the measures of quantum synchronization and correlations.
In Sec. IV, we investigate the influence of the temperature, coupling strength, and correlation time of the system on the
quantum synchronization and correlations of two qutrits.
In Sec. V, we summarize our results.

\section{Model and hierarchy equation method}
In this section, we introduce the model and hierarchy equation method used in the present work.
We consider a system formed by two qutrits with no direct interaction and the free Hamiltonian is (set $\hbar = 1$)
\begin{eqnarray}
H_S = \omega_1 J_1^z + \omega_2 J_2^z,
\end{eqnarray}
where $\omega_1$ and $\omega_2$ are frequencies of qutrit 1 and qutrit 2, respectively.
The detuning between two qutrits is $\Delta = \omega_2 - \omega_1$.
We assume two qutrits are put into a common thermal bath. The free Hamiltonian of the thermal bath is
\begin{eqnarray}
H_B = \sum_k \omega_k b_k^{\dag} b_k,
\end{eqnarray}
where $\omega_k$ is the frequency of the $k$th mode of the thermal bath.
The interaction Hamiltonian of two qutrits and bath is
\begin{eqnarray}
H_I = \sum_k g_k V(b_k^{\dag} + b_k),
\end{eqnarray}
where $g_k$ is the coupling strength between the qutrits and the $k$th mode of the bath.
Here, $b_k^{\dag}$ and $b_k$ are the creation and annihilation
operators of the thermal bath; V is the system operator coupled to the bath.
Without loss of generality, we suppose
\begin{eqnarray}
V = (1 + h)(J_1^z + J_2^z) + (1 - h)(J_1^x + J_2^x),
\end{eqnarray}
where $h$ is an anisotropy coefficient with $-1 \leq h \leq 1$.

In the interaction picture, the dynamics of the present system is \cite{22}
\begin{eqnarray}
\rho^I_S(t) &=& U(t) \rho_S(0), \label{sol} \\
U(t) &=& \mathcal{T} \exp\{ -\int _0^t dt_2\int _0^{t_2} dt_1 V(t_2)^{\times}[\Re[C(t_2 - t_1)]V(t_1)^{\times} \\ \nonumber
&& + i \Im[C(t_2 - t_1)] V(t_1)^\diamond] \} \rho_S(0),
\end{eqnarray}
where $\rho_S$ is the reduced density matrix of the system
and $\mathcal{T}$ is the chronological time-ordering operator.
Here, $O_1^{\times}O_2 \equiv [O_1,O_2] = O_1O_2 - O_2O_1$
and $O_1^{\diamond}O_2 \equiv \{O_1,O_2\} = O_1O_2 + O_2O_1$.
Note that $\Re[C(t_2 - t_1)]$ and $\Im[C(t_2 - t_1)]$ are the real
and imaginary parts of the bath time-correlation function
$C(t_2 - t_1) = \langle B(t_2)B(t_1)\rangle$, respectively, and
$B(t) = \sum_k (g_k b_k e^{-i\omega_k t} + g_k^*b^{\dag}e^{i\omega_k t})$.

In the present work, we choose the Drude-Lorentz spectrum \cite{21,22,23,24}
\begin{eqnarray}
J(\omega) = \omega \frac{2\lambda\gamma}{\pi(\gamma^2 + \omega^2)},
\end{eqnarray}
where $\lambda$ is the coupling strength between qutrits and bath,
$\gamma$ represents the width of the spectral distribution of the
bath mode. The quantity $1/\gamma$ represents the correlation time of the bath.
Particularly, if $\gamma$ is much larger than any other frequency scale, the Markovian approximation is valid.
For a bath with the Drude-Lorentz spectrum, the bath correlation function is \cite{23}
\begin{eqnarray}
\langle B(t_2)B(t_1)\rangle &=& \sum_{k=0}^{\infty} c_k e^{-\nu_k |t_2 - t_1|}, \label{bath_corr}\\
\nu_k &=& \frac{2\pi k }{\beta}(1 - \delta_{0k}) + \gamma \delta_{0k},\\
c_k &=& \frac{4\gamma \lambda \nu_k}{\beta(\nu_k^2 - \gamma^2)}(1 - \delta_{0k}) + \gamma \lambda [\cot(\frac{\gamma\beta}{2}) - i]\delta_{0k},
\end{eqnarray}
where $\beta = 1/(kT)$ is the inverse temperature of the thermal bath.
Using Eqs.(\ref{sol}) and (\ref{bath_corr}), the dynamics of the model can be describe by the hierarchy equation \cite{21}
\begin{eqnarray}
\dot{\rho}^n(t) &=& -(iH_s^{\times} + \sum_{\mu = 1,2}\sum_{k=0}^M n_{\mu k}\nu_k)\rho^n(t)\nonumber\\
&& - \sum_{\mu = 1,2} (\frac{2\lambda}{\beta\gamma} - i\lambda - \sum_{k=0}^M\frac{c_k}{\nu_k}) V^{\times}_{\mu}V^{\times}_{\mu}\rho^n(t) \nonumber\\
&& - i\sum_{\mu = 1,2}\sum_{k=0}^M n_{\mu k}[c_k V_{\mu}\rho^{n_{\mu k} ^-}(t) - c_k^* \rho^{n_{\mu k} ^-}(t)V_{\mu}]\nonumber\\
&& - i\sum_{\mu = 1,2}\sum_{k=0}^M V^{\times}_{\mu} \rho^{n_{\mu k} ^+}(t).
\end{eqnarray}
Note that $\rho^{n_{\mu k} ^+} = \rho^{n_{\mu k} \rightarrow n_{\mu k} + 1}$
($\rho^{n_{\mu k} ^-} = \rho^{n_{\mu k} \rightarrow n_{\mu k} - 1}$) denotes an increase (decrease) in the $\mu k$'th component of the multi-index.
It is worth noting that in the derivation of the above equation, the Markovian, Born, and rotating wave approximations are not used.
The hierarchy equation method is an exact method which is also suitable for strong- and ultrastrong-coupling systems.
The density matrix of two qutrits at arbitrary time can be obtained from the initial state of the system and
the above hierarchy equation of motion.
In the present work, we assume two qutrits are put into a common bath, i.e.,
$V_1 = V_2 = V$.

\section{Measures of quantum synchronization and correlations}
For a discrete-variable system, one can use the Husimi Q representation to
describe the phase portrait of a spin coherent state. In general, a spin coherent state is defined as \cite{18,19}
\begin{eqnarray}
|\theta, \phi\rangle = e^{-i\phi J_z} e^{-i\theta J_y} |J, J\rangle,
\end{eqnarray}
with the completeness relation
\begin{eqnarray}
\int_0^\pi d\theta \sin{\theta} \int_0^{2\pi} d\phi |\theta, \phi\rangle \langle \theta, \phi| = (4\pi)/(2J + 1).
\end{eqnarray}
For a spin 1 system, we have
\begin{eqnarray}
|\theta, \phi\rangle &=& \frac{e^{-i\phi}}{2}(1 + \cos\theta) |1,1\rangle + \frac{\sin{\theta}}{\sqrt{2}} |1,0\rangle \nonumber\\
&& + \frac{e^{i\phi}}{2} (1 - \cos{\theta}) |1,-1\rangle. \label{spin_coherent_states}
\end{eqnarray}

The measure of quantum synchronization proposed by Roulet and Bruder is defined as \cite{19}
\begin{eqnarray}
S_{r}(\phi) &=& \int_0^{2 \pi} d\phi_2\int_0^{\pi} d\theta_1 \int_0^{\pi} d\theta_2 \sin{\theta_1}
\sin{\theta_2} \nonumber\\
&& \times Q(\theta_1, \theta_2, \phi + \phi_2, \phi_2) - \frac{1}{2\pi}, \label{def_S}
\end{eqnarray}
where
\begin{eqnarray}
Q(\theta_1, \theta_2, \phi + \phi_2, \phi_2) &=& \frac{9}{16\pi^2} (\langle \theta_1, \phi + \phi_2|\otimes\langle \theta_2, \phi_2|) \nonumber\\
&& \rho (| \theta_1, \phi + \phi_2 \rangle \otimes | \theta_2, \phi_2\rangle). \label{Q}
\end{eqnarray}
Here, $Q(\theta_1, \theta_2, \phi + \phi_2, \phi_2)$ is the Husimi Q function and $\phi = \phi_1 - \phi_2$
is the relative phase of two spins. It can be viewed as a phase-space distribution
of density matrix $\rho$ based on spin coherent states. Note that $S_r(\phi)$ depends upon the relative phase $\phi$ explicitly.
Physically, it can be used to estimate whether two spins have tendency towards phase locking \cite{19}.
If $S_r(\phi)$ is always zero, then there is no fixed phase relation of two spins, i.e., no phase locking of two spins.
Using Eqs. (\ref{spin_coherent_states})-(\ref{Q}), we obtain the measure of synchronization of two spins 1 as
\begin{eqnarray}
S_r(\phi) &=& \frac{(32 \xi + 9\pi^2 \eta)}{256\pi}, \\
\xi &=& e^{2i\phi} \rho_{37} + e^{-2i\phi} \rho_{73}, \\
\eta &=& e^{i\phi}(\rho_{24} + \rho_{35} + \rho_{57} + \rho_{68}) \nonumber\\
&& + e^{-i\phi}(\rho_{42} + \rho_{53} + \rho_{75} + \rho_{86}),
\end{eqnarray}
where $\rho_{jk}$ is the element of density matrix $\rho$.

In order to measure the entanglement of two spins, we employ the logarithmic negativity which is defined by \cite{29,30}
\begin{equation}
E(\rho)\equiv \log_2{(1+2N)}= \log_2||\rho^{T}||,
\end{equation}
with $\rho^{T}$ being the partial transpose of density matrix $\rho$.
Here, $||\rho^{T}||$ is the trace norm of $\rho^{T}$ and $N$ is negativity defined by \cite{29,30}
\begin{equation}
N\equiv\frac{||\rho^{T}||-1}{2}.
\end{equation}
$N$ is the absolute value of the sum of the negative eigenvalues of $\rho^{T}$.

Now, we consider the quantum mutual information $I$ as a measure of all quantum correlations between two subsystems \cite{19,31}
\begin{eqnarray}
I = S(\rho_1) + S(\rho_2) - S(\rho),
\end{eqnarray}
with $\rho_1 = Tr_2(\rho)$ and $\rho_2 = Tr_1(\rho)$. Note that
$S(\rho) = - Tr[\rho \ln(\rho)]$ is the Von Neumann entropy of density matrix $\rho$.
In Ref. \cite{31}, the authors have proposed mutual information as an order parameter for quantum synchronization of a quantum system.

\section{Discussions}
\subsection{Influence of coupling strength $\lambda$}
\begin{figure}[tbp]
\centering {\scalebox{0.3}[0.3]{\includegraphics{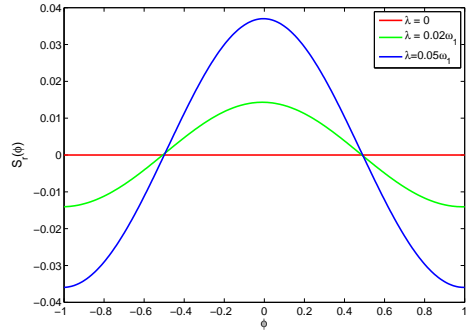}}}
\vspace*{8pt}
\caption{$S_r(\phi)$ is plotted as a function of $\phi$ for $\lambda = 0$ (red line), $\lambda = 0.02 \omega_1$ (green line),
and $\lambda = 0.05 \omega_1$ (blue line). The parameters are $\beta = 0.3/\omega_1, \gamma = 2\omega_1, \Delta = 0.01\omega_1$, and $h = -1$.
} \label{fig1}
\end{figure}

In Fig. 1, we plot $S_r(\phi)$ of steady state as a function of the relative phase $\phi$ of two spins for
different values of coupling strength $\lambda$. From Fig. 1, one can find that if the coupling constant is
zero, then $S_r(\phi)$ is always zero and there is no fixed phase relation between two spins. This implies
two spins can not be synchronized in the case of $\lambda = 0$. Physically, in the case of $\lambda = 0$,
there is no direct or indirect interaction between two qutrits. It is obvious that two qutrits can not be synchronized
without any interaction. On the other hand, the maximal value of $S_r(\phi)$ increases with the increase
of the coupling strength $\lambda$. For example, the maximum of $S_r(\phi)$
can be about 0.037 if $\lambda = 0.05$.
Therefore, two spins can be synchronized in the presence of the interaction of spins and the common bath.

\subsection{Influence of anisotropy coefficient $h$}

\begin{figure}[tbp]
\centering{\scalebox{0.3}[0.3]{\includegraphics{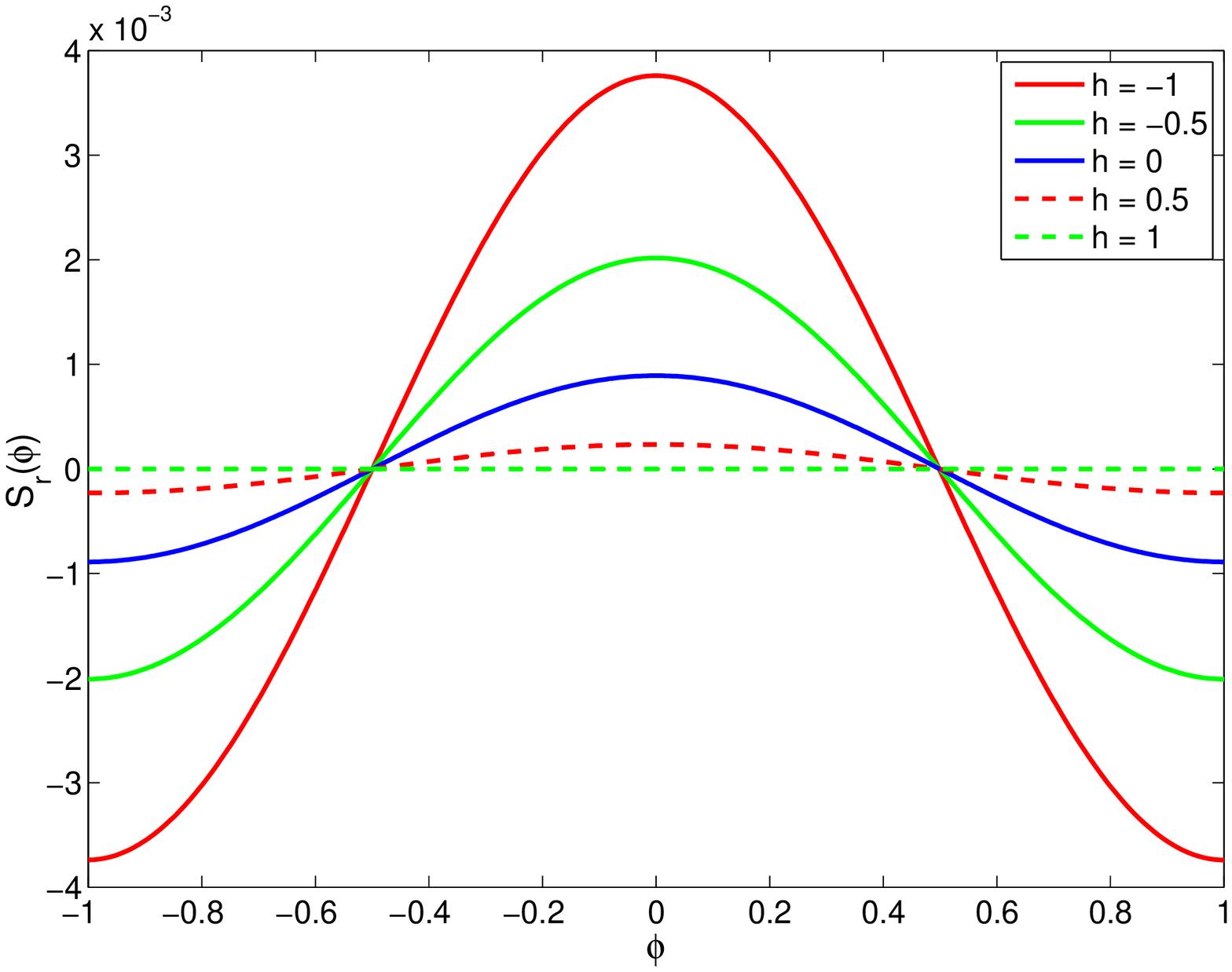}}}
\centering{\scalebox{0.3}[0.3]{\includegraphics{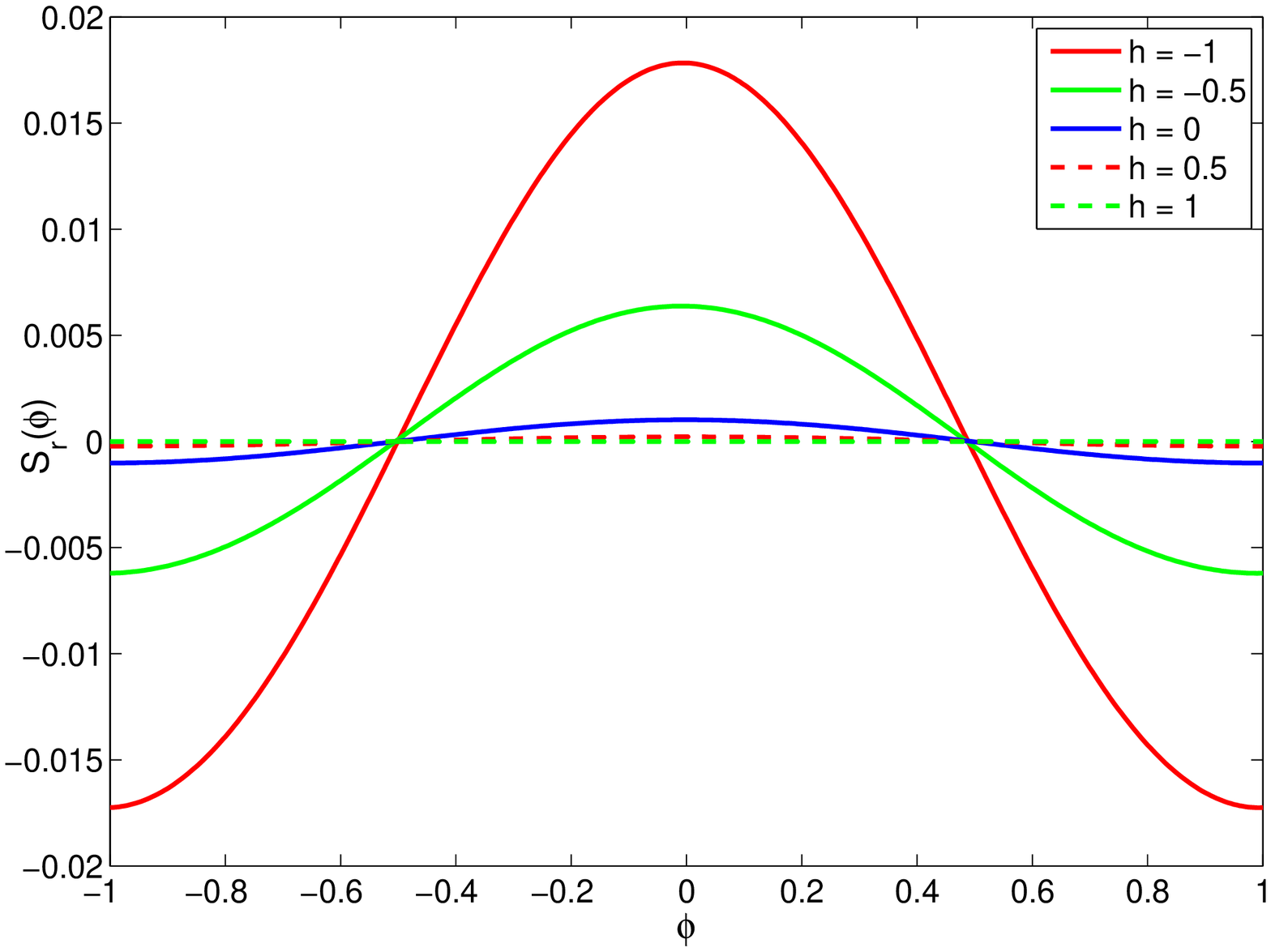}}}
\caption{$S_r(\phi)$ is plotted as a function of $\phi$ for
$\gamma = 0.2\omega_1$ (upper panel) and $\gamma = 4\omega_1$ (lower panel).
The parameters are $\beta = 0.3/\omega_1, \Delta = 0.01\omega_1$, and $\lambda = 0.03 \omega_1$.
} \label{fig2}
\end{figure}
We now turn to discuss the influence of the anisotropy coefficient $h$ on the synchronization of two spins.
The synchronization of two qubits within a common Markovian
environment has been investigated by employing the Bloch-Redfield master equation \cite{17}.
It is found that two qubits can not be synchronized for purely dephasing case. The Markovian
and Born approximation were employed in this work \cite{17}. In the following, we will show that
two spins can not be synchronized in purely dephasing case without using the Markovian and Born approximation.

In Fig. 2, we plot $S_r(\phi)$ as a function of $\phi$ for different values of $h$ with
$\gamma = 0.2\omega_1$ (upper panel) and $\gamma = 4\omega_1$ (lower panel).
One can clearly see that the maximal value of $S_r(\phi)$ decreases with the increase of the parameter $h$.
In particular, the values of $S_r(\phi)$ for $\gamma = 0.2 \omega_1$ (upper panel) and $\gamma = 4 \omega_1$ (lower panel)
are always zero if $h = 1$ and two spins can not be synchronized for the purely dephasing case. Note that, in Ref. \cite{17},
the authors have assumed that $\gamma \gg \omega_1$ and $\gamma \gg \omega_2$ in order to ensure the validity the Markovian approximation.
However, in the present work, we use the hierarchy equation method to investigate the present system
without the Markovian and Born approximations. More precisely, it is not necessary to assume $\gamma \gg \omega_1$
and $\gamma \gg \omega_2$ in our work. We extend the result of Ref. \cite{17}
to the case of non-Markovian bath, i.e., two spins without direct interaction can not be synchronized in the purely dephasing case.
We find dissipation is indispensable for the synchronization of two spins in Markovian or non-Markovian environment.

\subsection{Influence of temperature}
\begin{figure}[tbp]
\centering{\scalebox{0.3}[0.3]{\includegraphics{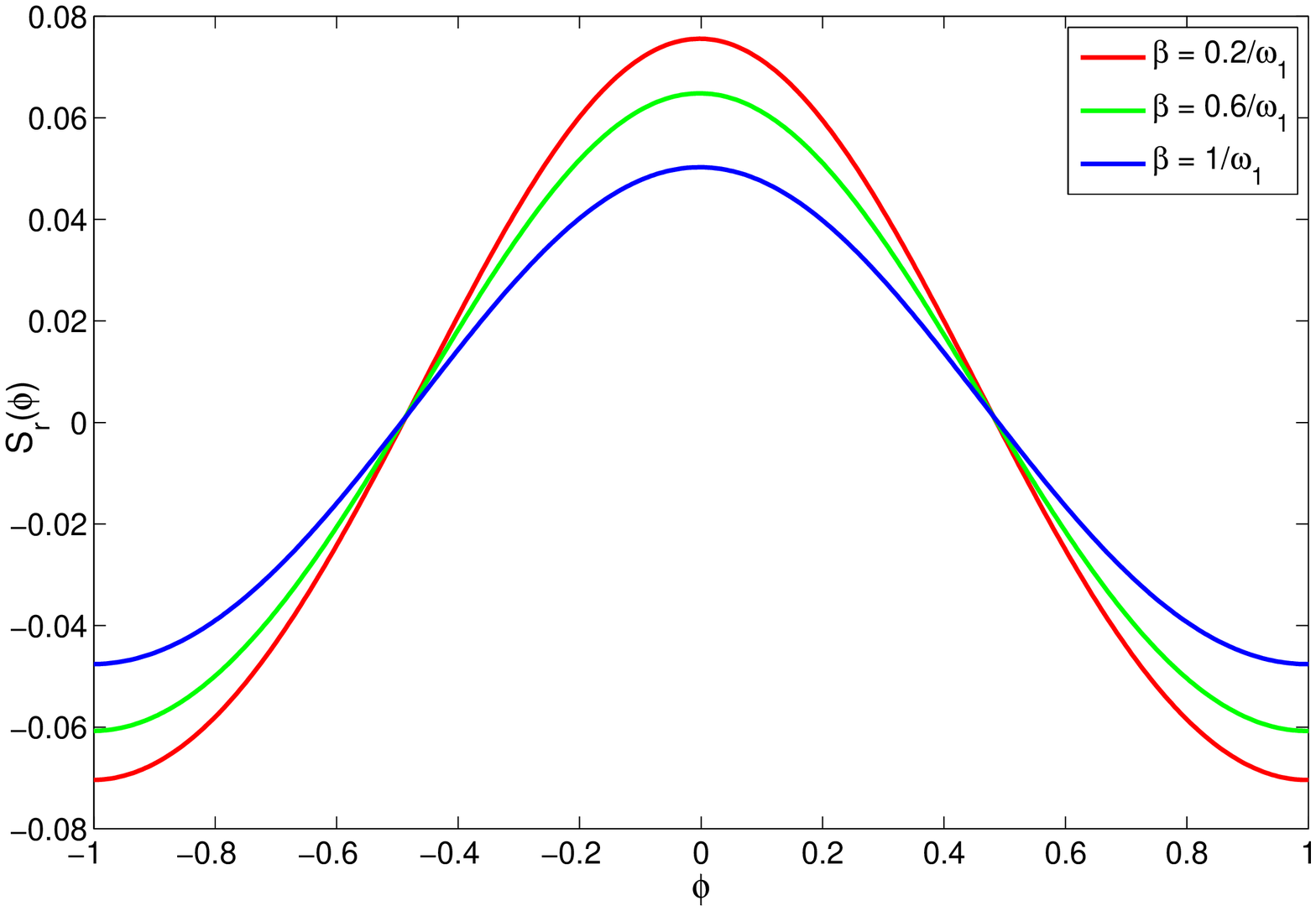}}}
\centering{\scalebox{0.3}[0.3]{\includegraphics{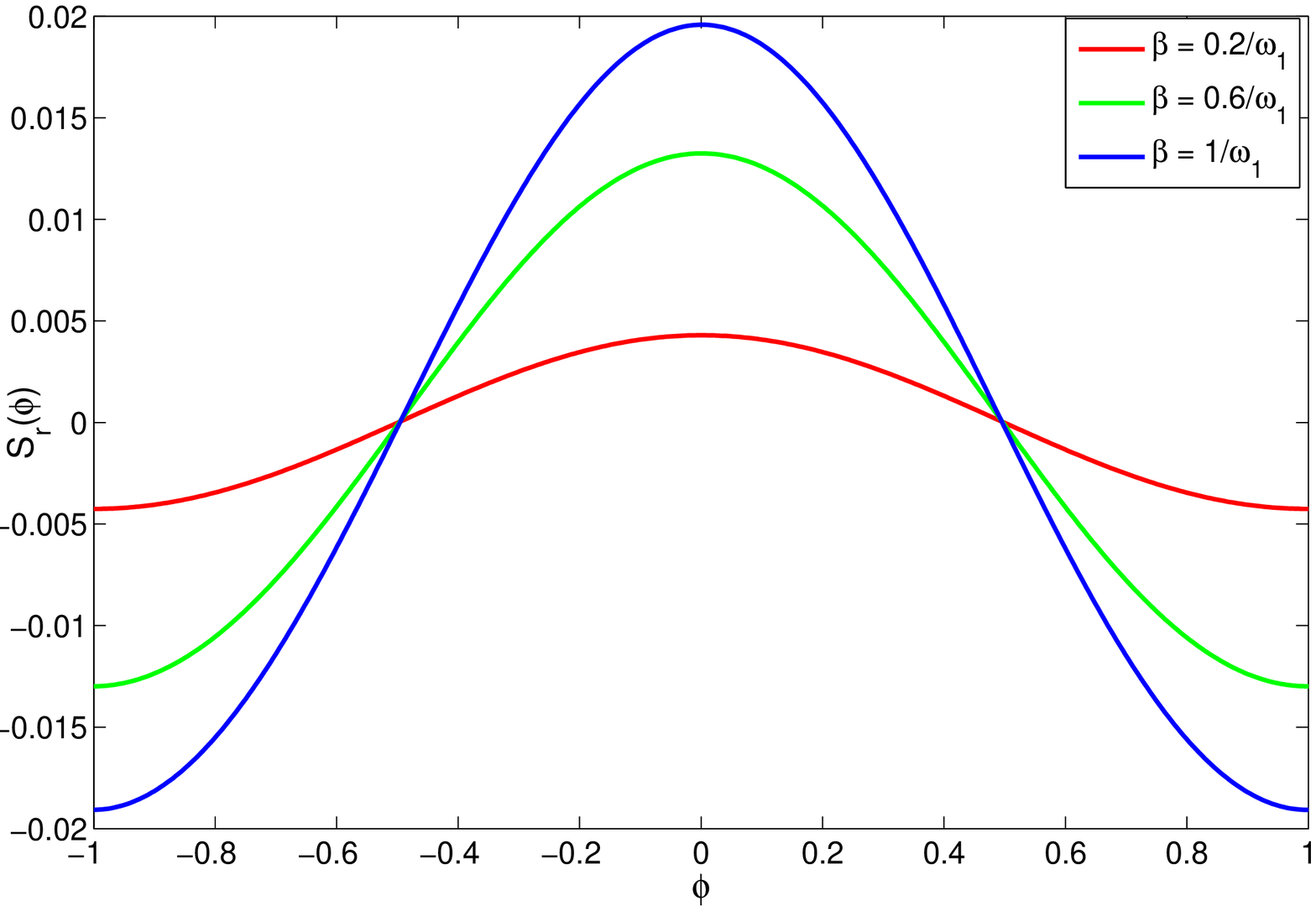}}}
\caption{$S_r(\phi)$ is plotted as a function of $\phi$ with
$\Delta = 0.001\omega_1$ (upper panel) and $\Delta = 0.1\omega_1$ (lower panel).
The parameters are $\gamma = 0.2 \omega_1$, $\lambda = 0.05 \omega_1$, and $h = -1$.
} \label{fig3}
\end{figure}

\begin{figure}[tbp]
\centering{\scalebox{0.3}[0.3]{\includegraphics{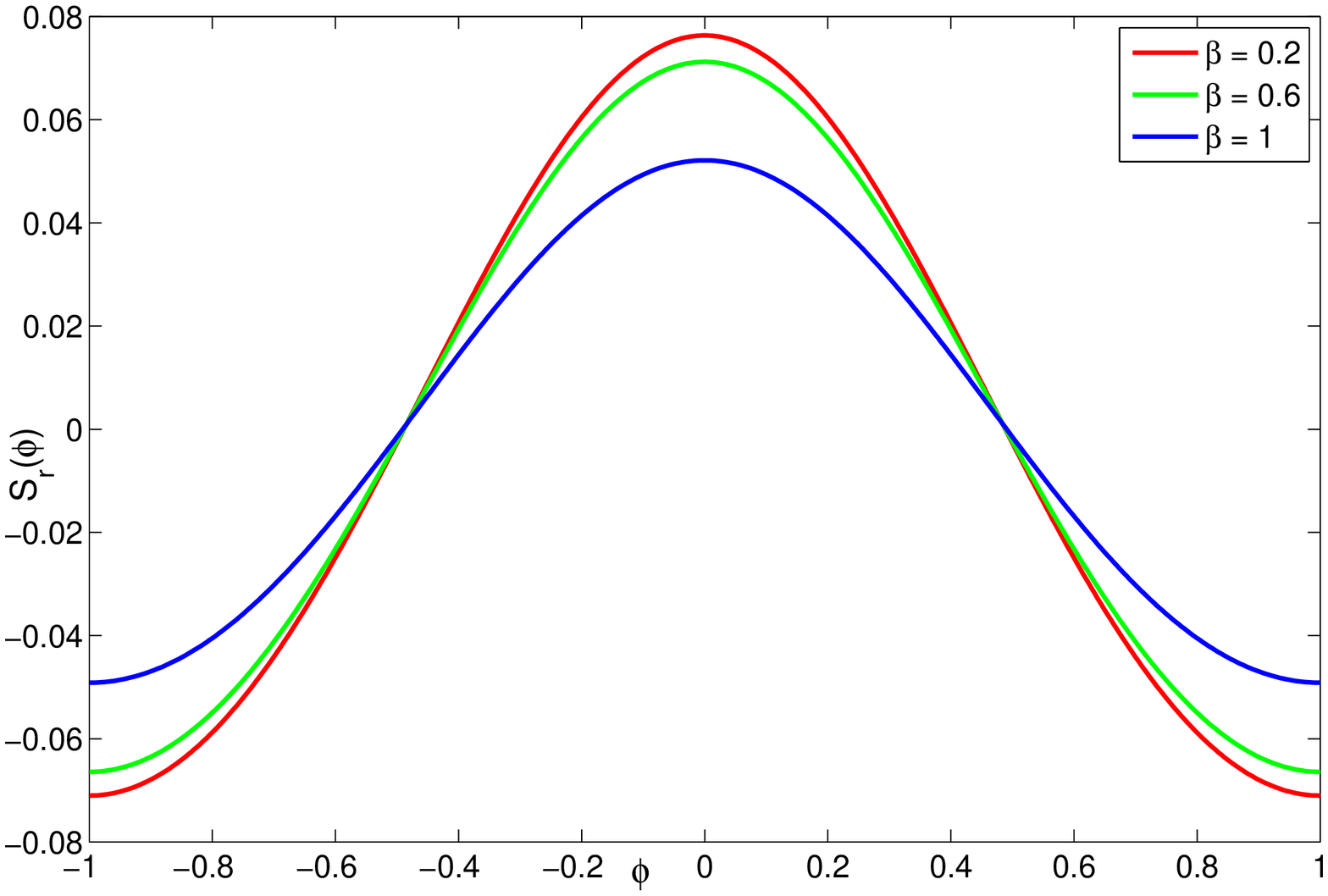}}}
\centering{\scalebox{0.3}[0.3]{\includegraphics{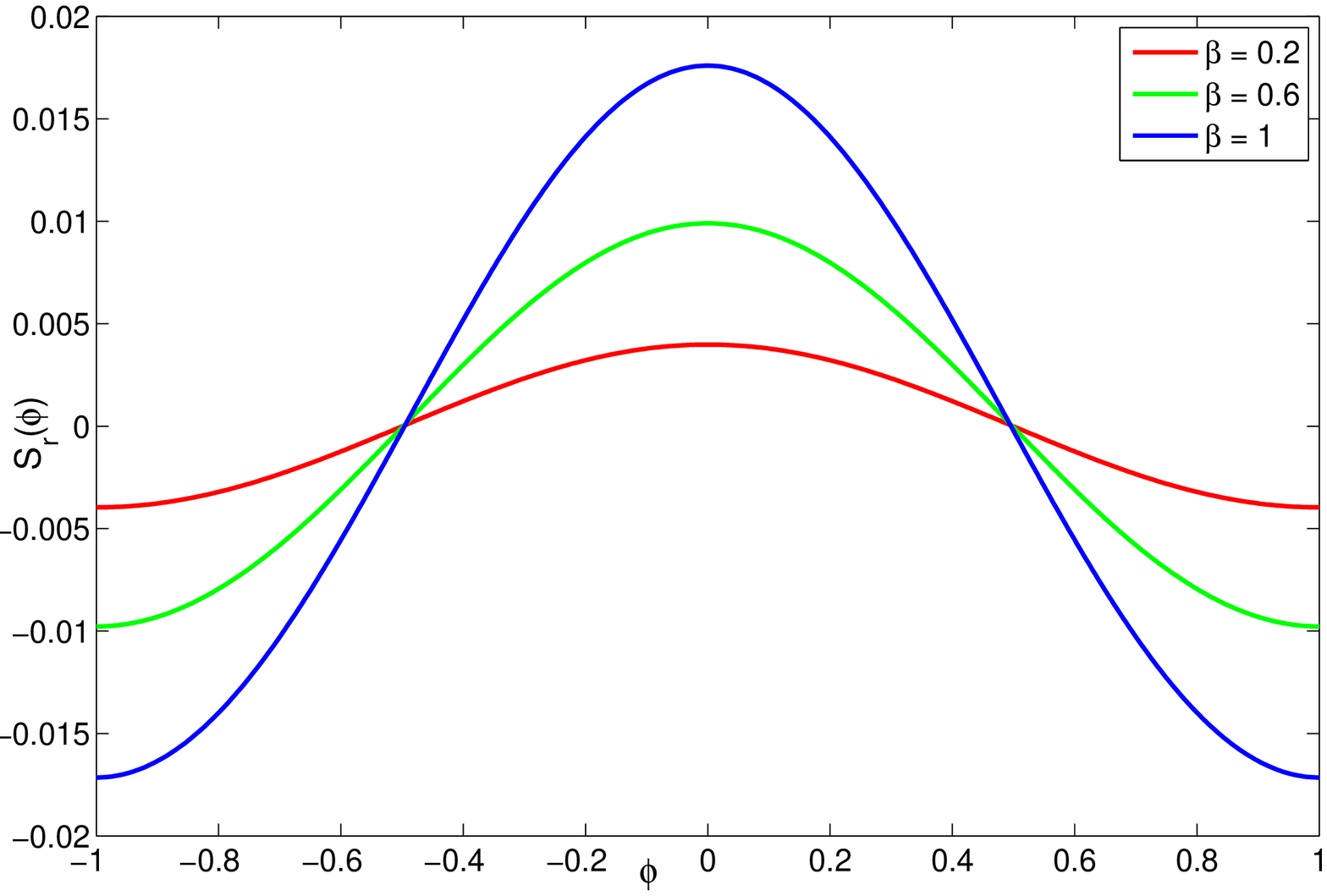}}}
\caption{$S_r(\phi)$ is plotted as a function of $\phi$ with
$\Delta = 0.001\omega_1$ (upper panel) and $\Delta = 0.1\omega_1$ (lower panel).
The parameters are $\gamma = 20 \omega_1$, $\lambda = 0.05 \omega_1$, and $h = -1$.
} \label{fig4}
\end{figure}

The synchronization of two spins has been studied with the help
of the Lindblad master equation and the temperature of the baths was assumed to zero \cite{18,19}.
In this section, we investigate the influence of the temperature of the bath.
Comparing the upper panel and lower panel of Fig. 3 ($\gamma = 0.2\omega_1$) and Fig. 4 ($\gamma = 20\omega_1$), we see the effects of the temperature of the bath
depends crucially on the detuning of two spins. On the one hand, if the detuning is much smaller than
the frequencies of spins ($\Delta \ll \omega_i$), the maximal value of $S_r(\phi)$ increases with the
increase of the temperature as one can see from the upper panel of Fig. 3 and Fig. 4.
On the other hand, the maximum of $S_r(\phi)$ decreases with the increase of the temperature
if $\Delta = 0.1\omega_1$ as one can see from the lower panel of Fig. 3 and Fig. 4.

One possible reason for the different influences of the temperature of the bath on $S_r(\phi)$ for
different detuning $\Delta$ is as follows. The interactions between the qutrits and the common bath plays an important role in
the generation of $S_r(\phi)$. The two qutrits interact with each other indirectly
via their direct interactions with the common bath. The temperature of the common bath plays a constructive role in this process. 
However, as the system evolves, the interactions
between the common bath and two qutrits can disturb the dynamics of two qutrits. In this case, the temperature of the bath plays a destructive role.
The steady state value of the quantum synchronization measure is a result of the two effects of the common bath.
If the detuning $\Delta$ is very small, the two qutrits can be synchronized
in a short time and the temperature of the bath plays a constructive role.
However, if the detuning $\Delta$ is large enough, it takes a long time to
synchronize the two qutrits and the temperature of the bath plays a destructive role.

\subsection{Arnold tongue}

\begin{figure}[tbp]
\centering{\scalebox{0.4}[0.3]{\includegraphics{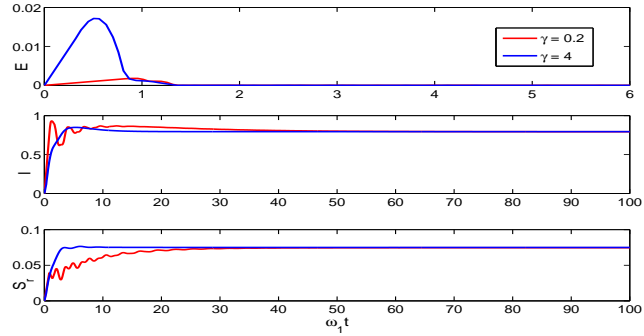}}}
\caption{The logarithmic
negativity $E$, mutual information $I$, and $S_r(\phi=0)$ are plotted
as functions of the dimensionless time $\omega_1 t$ for $\Delta = 0.001\omega_1$, $\lambda = 0.05\omega_1$, $\beta = 0.3/ \omega_1$, and $h = -1$.
} \label{fig5}
\end{figure}

\begin{figure}[tbp]
\centering{\scalebox{0.4}[0.3]{\includegraphics{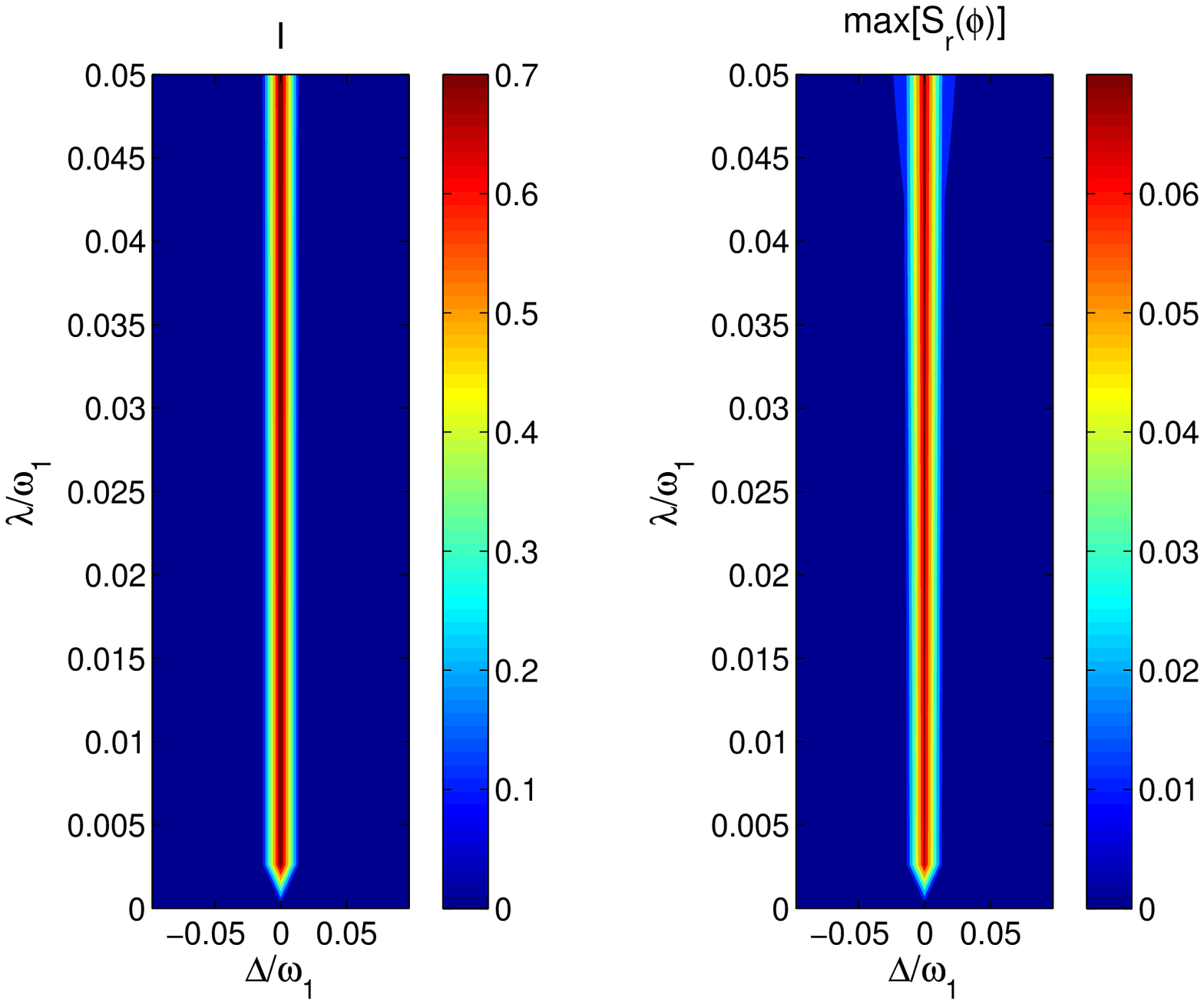}}}
\caption{The Arnold tongue of the present system. The quantum mutual information $I$ (left panel) and maximal value of $S_r(\phi)$ (right panel) are plotted
as functions of the detuning $\Delta$ and coupling strength $\lambda$ with
$\gamma = 0.2 \omega_1$, $\beta = 0.3/ \omega_1$, and $h = -1$.
} \label{fig6}
\end{figure}

\begin{figure}[tbp]
\centering{\scalebox{0.4}[0.3]{\includegraphics{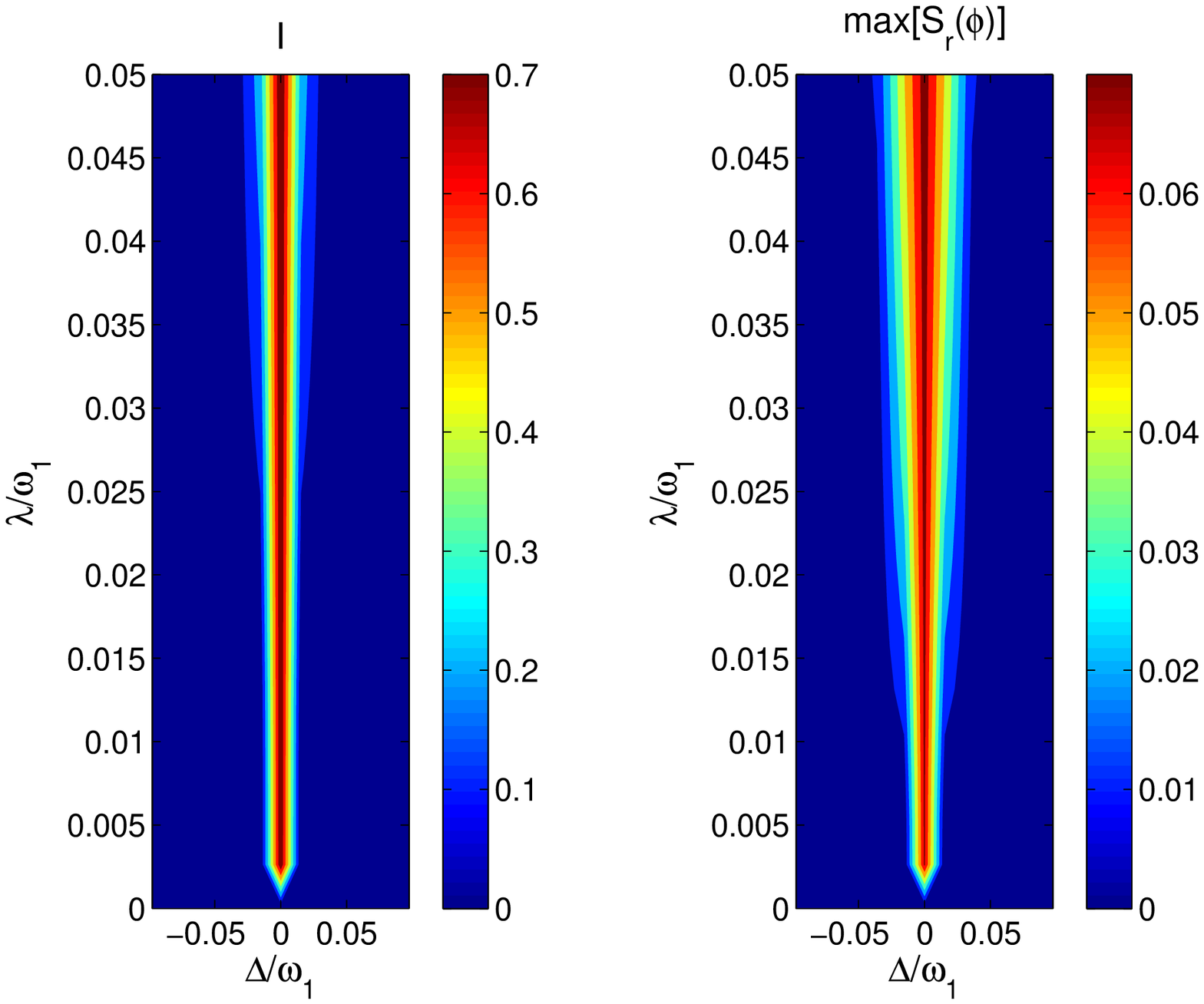}}}
\caption{The Arnold tongue of the present system. The quantum mutual information $I$ (left panel) and maximal value of $S_r(\phi)$ (right panel) are plotted
as functions of the detuning $\Delta$ and coupling strength $\lambda$ with
$\gamma = 4 \omega_1$, $\beta = 0.3/ \omega_1$, and $h = -1$.
} \label{fig7}
\end{figure}

In Fig. 5, we plot the logarithmic
negativity $E$, mutual information $I$, and $S_r(\phi=0)$
as functions of the dimensionless time $\omega_1 t$. The entanglement first increases and then decreases with time. 
Eventually, the entanglement becomes zero at $\omega_1 t \approx 1.5$ while
$I$ and $S_r$ are not zero at this time. After a certain time interval, the values of
$I$ and $S_r$ are not changed with time and the two spins are synchronized.

In order to see the steady state mutual information and synchronization measure more clearly, we plot the quantum mutual 
information $I$ (left panel) and maximum of $S_r(\phi)$ (right panel) as functions of
the detuning $\Delta$ and coupling strength $\lambda$ in Figs 6 and 7. The Arnold tongue which is the characteristic
property of synchronization can be observed in these figures. We calculate the logarithmic negativity
of two spins for many different parameters and find that there is no steady state entanglement even
in the presence of synchronization. This result is similar to the previous works \cite{3,31}.
Consequently, the mutual information has bee proposed as an order parameter for quantum synchronization \cite{31}.
 In the present work, we assume there is no direct interaction between two spins
and find they could not be entangled in the steady state, i.e., $E(\rho_{steady}) = 0$. However, the mutual information of two spins
at steady state could be larger than zero. Therefore, we plot the
mutual information of two spins in Figs. 6 and 7.
Comparing Fig. 6 and Fig. 7, we find the Arnold tongue could be adjusted by the parameter $\gamma$.
Particularly, the Arnold tongue in Fig. 6 is very narrow and it is usually very difficult observe
synchronization of two spins experimentally \cite{1}. If we increase the parameter $\gamma$,
then the Arnold tongue could be broadened significantly as one can see from Fig. 7.
Therefore, the synchronization of two spins could be observed in experiments more easily
if we increase the parameter $\gamma$.

\section{Conclusions}
In the present work, we have studied the quantum synchronization and correlations of two qutrits in one
non-Markovian environment with the help of the hierarchy equation method. There is no direct interaction
between two qutrits. Each qutrit interacts with the common non-Markovian bath. In order to measure quantum synchronization of discrete systems, we
adopted the measure $S_r(\phi)$ proposed by Roulet and Bruder \cite{18,19}.
This measure is based on the Husimi Q representation and spin coherent states.
We have investigated the influence of the temperature, correlation time, and coupling strength between qutrits and bath
on the quantum synchronzation and correlations of two qutrits without using the Markovian,
Born, and rotating wave approximations. The influence of dissipation and dephasing on the synchronization of two qutrits was also discussed.

We first discussed the influence of the coupling strength of qutrits and bath on the
quantum synchronization of two qutrits. If there is no interaction between each qutrit and
the common bath, then they do not interact with each other at all. Obviously,
they can not be synchronized in this case. If we increase the coupling strength of qutrits and bath,
they can be synchronized when dissipation is taken into accounted. Particularly, we found
that two spins without direct interaction in a non-Markovian bath can not be synchronized for purely
dephasing case which is a generalization of the Markovian case \cite{17}. In other words, dissipation is
indispensable for the quantum synchronization of two spins in non-Markovian or Markovian bath.

Then, we studied the influence of the temperature of the common bath on the quantum synchronization of two spins.
Our results show that the influence of the temperature of the common bath depends heavily on the detuning between two spins. If the detuning is much smaller
than the frequencies of two spins, the maximal value of $S_r(\phi)$ increases with the increase of the temperature.
However, when the detuning is not much smaller than the frequencies of two spins, the maximal value of $S_r(\phi)$ decreases with the increase of the temperature.

Finally, we plot the maximal value of $S_r(\phi)$ as a function of
the detuning $\Delta$ and coupling strength $\lambda$. The Arnold tongue which is the characteristic property of synchronization can be observed in the present model.
The logarithmic negativity of two spins for many different parameters was also calculated.
We find that there is no steady state entanglement even in the presence of synchronization \cite{3,31}. Therefore, we plot the
mutual information of two spins. The Arnold tongue could be adjusted by the parameter $\gamma$ significantly.
Particularly, the Arnold tongue is very narrow in the non-Markovian case $\gamma < \omega_i$ ($i = 1, 2$).
Thus, it is usually very difficult observe synchronization of two spins experimentally in the non-Markovian
case \cite{1}. If we increase the parameter $\gamma$, then the Arnold tongue could be broadened significantly.
Therefore, the synchronization of two spins could be observed in experiments more easily
if they are put into a Markovian environment.

\section*{Acknowledgments}
This work is supported by
the National Natural Science Foundation of China (Grant Nos. 11047115, 11365009 and 11065007),
the Scientific Research Foundation of Jiangxi (Grant Nos. 20122BAB212008 and 20151BAB202020.)

\end{document}